\documentclass[twocolumn, trackchanges]{aastex63}

\usepackage{graphicx}
\usepackage{apjfonts}
\usepackage{amsmath}
\usepackage[raggedright]{subfigure}
\usepackage{color}
\usepackage{hyperref}
\usepackage{acronym}

\newcommand{\psr}{PSR\,J1555$-$2908}
\def\Fermi{\textit{Fermi}}
\shorttitle{Does PSR\,J1555--2908 have another companion?} %%% < 44 characters
\shortauthors{\sc Nieder et al.}

\begin{document}

%% Acronyms
% no plural
\newacro{BIC}[BIC]{Bayesian Information Criterion}
\newacro{LAT}[LAT]{Large Area Telescope}
% normal plural (+s)
\newacro{MSP}[MSP]{millisecond pulsar}
\newacro{HTS}[HTS]{hierarchical triple system}

%% Title
\title{Is the black-widow pulsar PSR\,J1555--2908 in a hierarchical triple system?}

%% Authorlist
\author[0000-0002-5775-8977]{L.~Nieder}
\affiliation{Max-Planck-Institut f\"ur Gravitationsphysik (Albert-Einstein-Institut), 30167 Hannover, Germany}
\affiliation{Leibniz Universit\"at Hannover, 30167 Hannover, Germany}

\author[0000-0002-0893-4073]{M.~Kerr}
\affiliation{Space Science Division, Naval Research Laboratory, Washington, DC 20375-5352, USA}

\author[0000-0003-4355-3572]{C.~J.~Clark}
\affiliation{Jodrell Bank Centre for Astrophysics, Department of Physics and Astronomy, The University of Manchester, M13 9PL, UK}
\affiliation{Max-Planck-Institut f\"ur Gravitationsphysik (Albert-Einstein-Institut), 30167 Hannover, Germany}
\affiliation{Leibniz Universit\"at Hannover, 30167 Hannover, Germany}

\author[0000-0002-9032-7941]{P.~Bruel}
\affiliation{Laboratoire Leprince-Ringuet, CNRS, \'Ecole polytechnique, Institut Polytechnique de Paris, 91120 Palaiseau, France}

\author[0000-0002-6039-692X]{H.~T.~Cromartie}
\affiliation{Cornell Center for Astrophysics and Planetary Science, and Department of Astronomy, Cornell University, Ithaca, NY 14853, USA}
\affiliation{Hubble Fellowship Program Einstein Postdoctoral Fellow, USA}

\author[0000-0001-5799-9714]{S.~M.~Ransom}
\affiliation{National Radio Astronomy Observatory, 520 Edgemont Rd., Charlottesville, VA 22903, USA}

\author[0000-0002-5297-5278]{P.~S.~Ray}
\affiliation{Space Science Division, Naval Research Laboratory, Washington, DC 20375-5352, USA}

% corresponding author
\correspondingauthor{L.~Nieder}
\email{lars.nieder@aei.mpg.de}

% dates
\received{XXXX}
\revised{YYYY}
\accepted{ZZZZ}
\published{AAAA}

%%%%%%%%%%%%%%%%%%%%%%%%%%%%%%%%%%%%%%%%%%%
\begin{abstract} %%%<250words
%%%%%%%%%%%%%%%%%%%%%%%%%%%%%%%%%%%%%%%%%%%

\noindent
The $559$\,Hz black-widow pulsar \psr{}, originally discovered in radio, is also a bright gamma-ray pulsar.  Timing its pulsations using 12\,yr of \Fermi{}-LAT gamma-ray data reveals long-term variations in its spin frequency that are much larger than is observed from other millisecond pulsars.  While this variability in the pulsar rotation rate could be intrinsic ``timing noise'', here we consider an alternative explanation: the variations arise from the presence of a very-low-mass third object in a wide multi-year orbit around the neutron star and its low-mass companion.  With current data, this hierarchical-triple-system model describes the pulsar's rotation slightly more accurately than the best-fitting timing-noise model.  Future observations will show if this alternative explanation is correct.
\end{abstract}


%%%%%%%%%%%%%%%%%%%%%%%%%%%%%%%%%%%%%%%%%%%
\section{Introduction}\label{s:intro_j1555_HTS}
%%%%%%%%%%%%%%%%%%%%%%%%%%%%%%%%%%%%%%%%%%%

\psr{}, a neutron star spinning rapidly at $559$\,Hz, resides in a tight $5.5$\,hr binary system (Ray et al. 2022, submitted, hereafter Paper~I) with a low-mass ($0.06\,M_{\odot}$) companion (Kennedy et al. 2022, submitted). This \ac{MSP} was first detected in a Green Bank Telescope (GBT) pulsar survey which targeted steep-spectrum radio sources \citep{frail2018} within the localization region of \Fermi{}-\acl{LAT} \cite[\acsu{LAT};][]{atwood2009} gamma-ray sources.  After the radio discovery, gamma-ray pulsations were detected, allowing the timing measurement of the system parameters over the $12$-yr \Fermi{} mission time span.

Timing analysis using the \ac{LAT} data reveals variations of the spin frequency $f$ which are larger than is typical for \acp{MSP} \citep{kerr2015b}.  Such variations are often seen in young gamma-ray pulsars, where they are labeled as ``timing noise''.  Timing noise is also present in \ac{MSP}, but its weaker amplitude \citep{Shannon10} is typically not detectable with the \ac{LAT} (Kerr et al. 2022, submitted).  Instead, the intrinsic rotational phase is generally well described by a quadratic function of time, i.e. a constant spin-down rate $\dot{f}$, with no detectable noise components above the measurement uncertainty.  In contrast, for \psr{} four additional terms in the rotational phase model are required.\footnote{In a gamma-ray timing analysis, timing noise is easily distinctive from the variations of the orbital period which are often seen in ``spider'' pulsars. While both variations happen on multi-year timescales, spin-frequency variations manifest as pulse phase drift, and variations of the short ($\lesssim 1$\,d) orbital period smear the pulse profile out.}

Such strong variations suggest an alternative explanation.  For example, in the case of the millisecond pulsar PSR\,J1024$-$0719, long thought to be isolated, the measurement of additional higher-order spin-frequency derivatives led to the conclusion that the pulsar could be in a very-long-period ($>2$\,kyr) orbit with a low-mass companion \citep{Kaplan2016+WideBinary}.

In this paper, we discuss a similar alternative to the timing noise explanation for \psr{}: that the variations arise from the presence of a third body in the system.  This additional object is in a wide, multi-year orbit around the closely orbiting neutron star and its low-mass companion.  An extreme example of such a system would be the hierarchical-triple-system pulsar PSR\,J0337$+$1715 \citep{ransom2014}.

With the currently available data, this model describes the pulsar as well as the timing-noise model, but provides a simple and clear physical explanation for the spin frequency variations.

%%%%%%%%%%%%%%%%%%%%%%%%%%%%%%%%%%%%%%%%%%%
\section{Rotational phase model}\label{s:model}
%%%%%%%%%%%%%%%%%%%%%%%%%%%%%%%%%%%%%%%%%%%

To precisely track the rotational phase, the photon arrival times need to be corrected for the line-of-sight motion of the pulsar, if it is in an orbit with one or more companions.  For a simple \ac{HTS}, we assume that the gravitational interaction between the two companions can be neglected, and the third body orbits the center of mass of the inner binary system.  The photon times, corrected for the pulsar's motion around the barycenter of the triple system, $t_{\rm tb}$, can be expressed as a function of the photon's emission time, $t_{\rm em}$, as
%\begin{linenomath}
\begin{equation} \label{eq:hts_time}
	t_{\rm tb} = t_{\rm em} + x_{\rm i} \Delta_{\rm i}(\Omega_{\rm i} t_{\rm em}) + x_{\rm o} \Delta_{\rm o}(\Omega_{\rm o} t_{\rm em}) \,.
\end{equation}
%\end{linenomath}
Here, $x_{\rm i}$ and $x_{\rm o}$ are the times light needs to travel the semimajor axis of the pulsar's orbit due to the inner (i) and outer (o) companion projected onto the line of sight.  $\Delta_{\rm i}$ and $\Delta_{\rm o}$ are dimensionless functions that describe the respective modulations depending on the orbital phases, $\Omega_{\rm i} t_{\rm em}$ and $\Omega_{\rm o} t_{\rm em}$.  The orbital angular frequencies $\Omega_{\rm i}$ and $\Omega_{\rm o}$ are defined by the orbital periods $P_{\rm i}$ and $P_{\rm o}$ via $\Omega_{\rm i} = 2\pi / P_{\rm i}$ and $\Omega_{\rm o} = 2\pi / P_{\rm o}$.

Following \cite{edwards2006}, we can make $x_{\rm i} \Delta_{\rm i}(\Omega_{\rm i} t_{\rm em}) + x_{\rm o} \Delta_{\rm o}(\Omega_{\rm o} t_{\rm em})$ a function of $t_{\rm tb}$ by Taylor expansion.  To second order (i.e. including terms of order $(x_{\rm i}\Omega_{\rm i})^2$, $(x_{\rm o}\Omega_{\rm o})^2$, $x_{\rm i}x_{\rm o}\Omega_{\rm i}^2$, and $x_{\rm i}x_{\rm o}\Omega_{\rm o}^2$), we find
%\begin{linenomath}
\begin{align}
	x_{\rm i} &\Delta_{\rm i}(\Omega_{\rm i} t_{\rm em}) + x_{\rm o} \Delta_{\rm o}(\Omega_{\rm o} t_{\rm em}) \label{eq:hts_phasemodel}\\
	&\approx \vphantom{\frac{1}{2}} \left(x_{\rm i} \Delta_{\rm i}(\Omega_{\rm i} t_{\rm tb}) + x_{\rm o} \Delta_{\rm o}(\Omega_{\rm o} t_{\rm tb})\right) \nonumber\\
	& \quad \times \left[\vphantom{\frac{1}{2}} 1 - \left(x_{\rm i}\Omega_{\rm i} \Delta_{\rm i}^{\prime}(\Omega_{\rm i} t_{\rm tb}) + x_{\rm o}\Omega_{\rm o} \Delta_{\rm o}^{\prime}(\Omega_{\rm o} t_{\rm tb})\right)\right. \nonumber\\
	& \quad \quad \quad + \left(x_{\rm i}\Omega_{\rm i} \Delta_{\rm i}^{\prime}(\Omega_{\rm i} t_{\rm tb}) + x_{\rm o}\Omega_{\rm o} \Delta_{\rm o}^{\prime}(\Omega_{\rm o} t_{\rm tb})\right)^2 \vphantom{\frac{1}{2}} \nonumber\\
	& \quad \quad \quad + \frac{1}{2} \left(x_{\rm i}\Omega_{\rm i}^2 \Delta_{\rm i}^{\prime\prime}(\Omega_{\rm i} t_{\rm tb}) + x_{\rm o}\Omega_{\rm o}^2 \Delta_{\rm o}^{\prime\prime}(\Omega_{\rm o} t_{\rm tb})\right) \nonumber\\
	& \quad \quad \quad \quad \times \left. \left(x_{\rm i} \Delta_{\rm i}(\Omega_{\rm i} t_{\rm tb}) + x_{\rm o} \Delta_{\rm o}(\Omega_{\rm o} t_{\rm tb})\right) \vphantom{\frac{1}{2}} \right] \,, \nonumber
\end{align}
%\end{linenomath}
where $\Delta^{\prime}$ and $\Delta^{\prime\prime}$ denote the first and second derivative of $\Delta$ with respect to $\Omega t$.  For wide, multi-year outer orbits of low-mass companions, $x_{\rm o} \Omega_{\rm o} \ll x_{\rm i} \Omega_{\rm i}$, so we neglect terms multiplied by this quantity, i.e. $\Delta_{\rm o}^{\prime} = 0$ and $\Delta_{\rm o}^{\prime\prime} = 0$.

In the case of a circular orbit, $\Delta$ takes the simple form of a sinusoid, $\Delta(t) = \sin(\Omega(t - T_{\rm asc}))$, with time of ascending node $T_{\rm asc}$.  Hence, $\Delta^{\prime}(t) = \cos(\Omega(t - T_{\rm asc}))$ and $\Delta^{\prime\prime}(t) = -\sin(\Omega(t - T_{\rm asc}))$.  Eq.~\eqref{eq:hts_phasemodel} can also be applied to the widely used small-eccentricity-orbit ``ELL1 model'' \citep{lange2001}, or the larger-eccentricity models presented in \citet{Nieder2020+Methods}.

For the outer companion's orbit of \psr{}, we use the full ``BT model'', $\Delta(t) = \sin \omega (\cos E - e) + \sqrt{1-e^2} \cos\omega \sin E$ with eccentricity $e$, and angle of periastron $\omega$ \citep{blandford1976}.  $E$ is defined via $E - e \sin E = \Omega_{\rm b} (t - T_0)$ and is computed numerically.  Instead of using the parameter set $\{T_0, e, \omega\}$, which is degenerate for small eccentricities, we use $T_{\rm asc} = T_0 - \omega/\Omega_{\rm b}$, $\epsilon_1 = e \sin\omega$, and $\epsilon_2 = e \cos\omega$ \citep{lange2001}.

In Section~\ref{s:gamma}, we show that \psr{} can be well described as an \ac{HTS} with the inner orbit being circular and the outer orbit being eccentric.

For a timing analysis, it is helpful to have a good starting point in the multi-dimensional parameter space.  The parameters describing the potential outer orbit can be roughly estimated from the timing solution presented in Paper~I, which requires four additional frequency derivatives (``FX model'').

First, we need to derive how the observed frequency changes in the case of an outer circular orbit.  Assuming only a non-zero spin frequency $f$ and a non-zero first time-derivative $\dot{f}$, the additional evolution of the spin frequency over time can be written in two ways:
%\begin{linenomath}
\begin{align}
	f(t)& - f_* - \dot{f}_*(t-t_0) \nonumber\\
	&= \frac{1}{2} f^{(2)}_*(t-t_0)^2 + \frac{1}{6} f^{(3)}_*(t-t_0)^3 + \frac{1}{24} f^{(4)}_*(t-t_0)^4 \label{eq:f_x}\\
	& \quad \quad + \frac{1}{120} f^{(5)}_*(t-t_0)^5 \nonumber\\
	&= \Delta f_* + \Delta \dot{f}_*(t-t_0) \label{eq:f_hts} \\
	& \quad \quad - (f_* + \Delta f_*) x_{\rm o} \Omega_{\rm o} \cos\left(\Omega_{\rm o} (t - T_{\rm asc,o})\right) \,. \nonumber
\end{align}
%\end{linenomath}
Here, we denote the $n$th frequency derivative as $f^{(n)}$, and use an asterisk to indicate parameters measured with the FX model, and $\Delta f_* = f - f_*$ as well as $\Delta \dot{f}_* = \dot{f} - \dot{f}_*$. $t_0$ denotes the chosen reference time.

Second, since all parameters in Eq.~\eqref{eq:f_x} are measured, we fit Eq.~\eqref{eq:f_hts} to Eq.~\eqref{eq:f_x} over the valid time span of the pulsar timing solution.  To do this, we used \texttt{curve\_fit} from \texttt{scipy}.  This gives initial values for the five free parameters $\{\Delta f_*, \Delta \dot{f}_*,x_{\rm o},\Omega_{\rm o}, T_{\rm asc,o}\}$ from Eq.~\eqref{eq:f_hts}.

From the fitted values, we find the inner binary barycenter's potential movement would have a radius of a few tens of kilometers with an orbital period of $\sim 10$\,yr, indicating a low-mass companion in a wide orbit.

%%%%%%%%%%%%%%%%%%%%%%%%%%%%%%%%%%%%%%%%%%%
\section{Gamma-ray timing analysis}\label{s:gamma}
%%%%%%%%%%%%%%%%%%%%%%%%%%%%%%%%%%%%%%%%%%%

In our timing analysis, we use the \ac{LAT} data prepared for the timing in Paper~I.  These data included P8R3 \texttt{SOURCE}-class photons \citep{atwood2013,Bruel2018+P305} detected by \ac{LAT} between 2008 August 03 and 2020 August 05 within a $5\deg$ region of interest (RoI) around the pulsar position, with energies greater than $100\,$MeV, and with a maximum zenith angle of $90\deg$.

The timing analysis is done twice, very similarly to the approach in Paper~I.  Firstly, we refit all the parameters of the FX model, i.e. we did not fix any parameters to the radio timing solution.  Secondly, we amended the timing code with the \ac{HTS} model given in Equations \eqref{eq:hts_time} and \eqref{eq:hts_phasemodel}.  To judge the timing results, we used the $H$ statistic \citep{dejager1989,kerr2011} with the photon probability weights optimized following \cite{Bruel2019+Weights}, the likelihood $\mathcal{L}$ \citep{abdo2013}, and the \acl{BIC} \citep[\acsu{BIC};][]{schwarz1978}.  The latter is based on the likelihood but with a penalty factor proportional to the degrees of freedom to penalize higher-dimensional models.

In the timing analysis, we tried to find the parameters that maximize $\mathcal{L}$ for a given model.  In the case of the \ac{HTS} model, we tested the circular and eccentric options, where the latter requires two additional parameters.  The eccentric orbit model is favored by all three statistics.  For the FX model, we added spin-frequency derivatives until the test statistics $H$ and $\mathcal{L}$ stopped improving significantly, which is in turn disfavored by the \ac{BIC}.  Here, adding six additional derivatives in principle leads to $H$ and $\mathcal{L}$ values of a similar size to the \ac{HTS} model.  However, the last model favored by the \ac{BIC} is the one with four additional frequency derivatives.  The final timing solutions for the FX and \ac{HTS} models are reported in Table~\ref{t:timing_j1555_HTS}.

In the timing analysis with the \ac{HTS} model, the highest test statistics are found for an eccentric ($e_{\rm o} \sim 0.27$) outer orbit of $\sim 4{,}500$\,d.  With the \ac{LAT} data span being only slightly shorter, the spin parameters $f$ and $\dot{f}$ are still correlated with the outer-orbit Keplerian parameters $P_{\rm o}$, $x_{\rm o}$ and $T_{\rm asc, o}$.  However, this covariance is small enough to not pose a problem for the timing analysis.  Note, that the timing analysis does not give any information on the inclination between the orbits as only the line-of-sight motion can be measured.

The highest test statistics, $H$ and $\mathcal{L}$, seem to favor the \ac{HTS} model over the FX model (see Table~\ref{t:timing_j1555_HTS}).  Since each model has a different number of free parameters, we invoke the \ac{BIC}.  The \ac{BIC}'s penalty for the additional parameter removes most of the advantage the \ac{HTS} model had. However, the \ac{BIC} still slightly favors the \ac{HTS} model over the FX model ($\Delta\textrm{BIC} = -1.6$).  Our final timing solutions over $12$ years of \ac{LAT} data using the FX model and the \ac{HTS} model are shown in Table~\ref{t:timing_j1555_HTS}.

In order to more directly compare the timing noise to that of other \acp{MSP}, we have analyzed it by following the methods of \citet{Lentati14}, adapted to gamma-ray data as in Kerr et al. (2022, submitted).  In this approach, the timing noise is assumed to originate in a stationary process, thus having a unique description with a power spectral density which we assume to follow a power law, $P(f)=P_{\mathrm{tn}}(f\times\mathrm{yr})^{-\Gamma_{\mathrm{tn}}}$.  The timing model parameters, timing noise parameters $\Gamma$ and $P_{\mathrm{tn}}$, and the Fourier coefficients of the timing noise process are all determined jointly by maximizing the likelihood, which includes a Gaussian component that constrains the Fourier coefficients to follow the specified power law.  We estimate uncertainties on the timing noise parameters from the curvature of the likelihood at its maximum value, and obtain $\log_{10}(P_{\mathrm{tn}}\times\mathrm{s^{-2}\,yr})=-13.3\pm1.3$ and spectral index $\Gamma_{\mathrm{tn}}=6.0\pm1.4$.

%------------------------------------------------------------------------------
\begin{deluxetable*}{lcc}
\tablewidth{0.99\columnwidth}
\tablecaption{\label{t:timing_j1555_HTS} Timing solutions for \psr{} with FX and \ac{HTS} models.}
\tablecolumns{3}
\tablehead{ Parameter & FX value & \ac{HTS} value }
\startdata
Range of observational data (MJD) & \multicolumn{2}{c}{$54681$ -- $59066$} \\
Reference epoch (MJD)             & \multicolumn{2}{c}{$57800.0$} \\
\cutinhead{Statistics}
$H$ statistic
& $908.7$
& $940.2$ \\
Log-Likelihood $\log \mathcal{L}$
& $283.7$
& $288.1$ \\
\ac{BIC}
& $-444.9$
& $-446.5$ \\
\cutinhead{Timing parameters}
R.A., $\alpha$ (J2000.0)
& $15^{\rm h}55^{\rm m}40\fs6587(2)$
& $15^{\rm h}55^{\rm m}40\fs6586(2)$ \\
Decl., $\delta$ (J2000.0)
& $-29\arcdeg08\arcmin28\farcs424(8)$
& $-29\arcdeg08\arcmin28\farcs424(7)$ \\
Spin frequency, $f$ (Hz)
& $559.44000642607(5)$
& $559.44000642558(17)$ \\
1st spin-frequency derivative, $\dot{f}$ (Hz s$^{-1}$)
& $-1.39361(18)\times 10^{-14}$
& $-1.39232(12)\times 10^{-14}$ \\
2nd spin-frequency derivative, $f^{(2)}$ (Hz s$^{-2}$)
& $5.6(4)\times 10^{-25}$ \\
3rd spin-frequency derivative, $f^{(3)}$ (Hz s$^{-3}$)
& $1(2)\times 10^{-33}$ \\
4th spin-frequency derivative, $f^{(4)}$ (Hz s$^{-4}$)
& $-2.0(6)\times 10^{-40}$ \\
5th spin-frequency derivative, $f^{(5)}$ (Hz s$^{-5}$)
& $-3.4(8)\times 10^{-48}$ \\
\cutinhead{Inner-orbit binary parameters}
Proj. semimajor axis, $x_{\rm i}$ (lt-s)
& $0.151445(3)$
& $0.151444(3)$ \\
Orbital period, $P_{\rm i}$ (days)
& $0.23350026827(13)$
& $0.23350026831(12)$ \\
Epoch of ascending node, $T_{\rm asc,i}$ (MJD)
& $57785.5393610(8)$
& $57785.5393612(8)$ \\
\cutinhead{Outer-orbit binary parameters}
Proj. semimajor axis, $x_{\rm o}$ (lt-s) &
& $0.000155(20)$ \\
Orbital period, $P_{\rm o}$ (days) &
& $4.5(4)\times 10^{3}$ \\
Epoch of ascending node, $T_{\rm asc,o}$ (MJD) &
& $54840(140)$ \\
1st Laplace-Lagrange parameter, $\epsilon_{1,\rm{o}}$ &
& $0.27(5)$ \\
2nd Laplace-Lagrange parameter, $\epsilon_{2,\rm{o}}$ &
& $0.02(7)$ \\
\enddata
\tablecomments{Numbers in parentheses are statistical $1\sigma$ uncertainties on the final digits.  The JPL DE405 solar system ephemeris has been used, and times refer to TDB.}
\end{deluxetable*}
%------------------------------------------------------------------------------

%%%%%%%%%%%%%%%%%%%%%%%%%%%%%%%%%%%%%%%%%%%
\section{Discussion}\label{s:discussion}
%%%%%%%%%%%%%%%%%%%%%%%%%%%%%%%%%%%%%%%%%%%

\begin{figure}[t]
	\centering
	\includegraphics[width=\columnwidth]{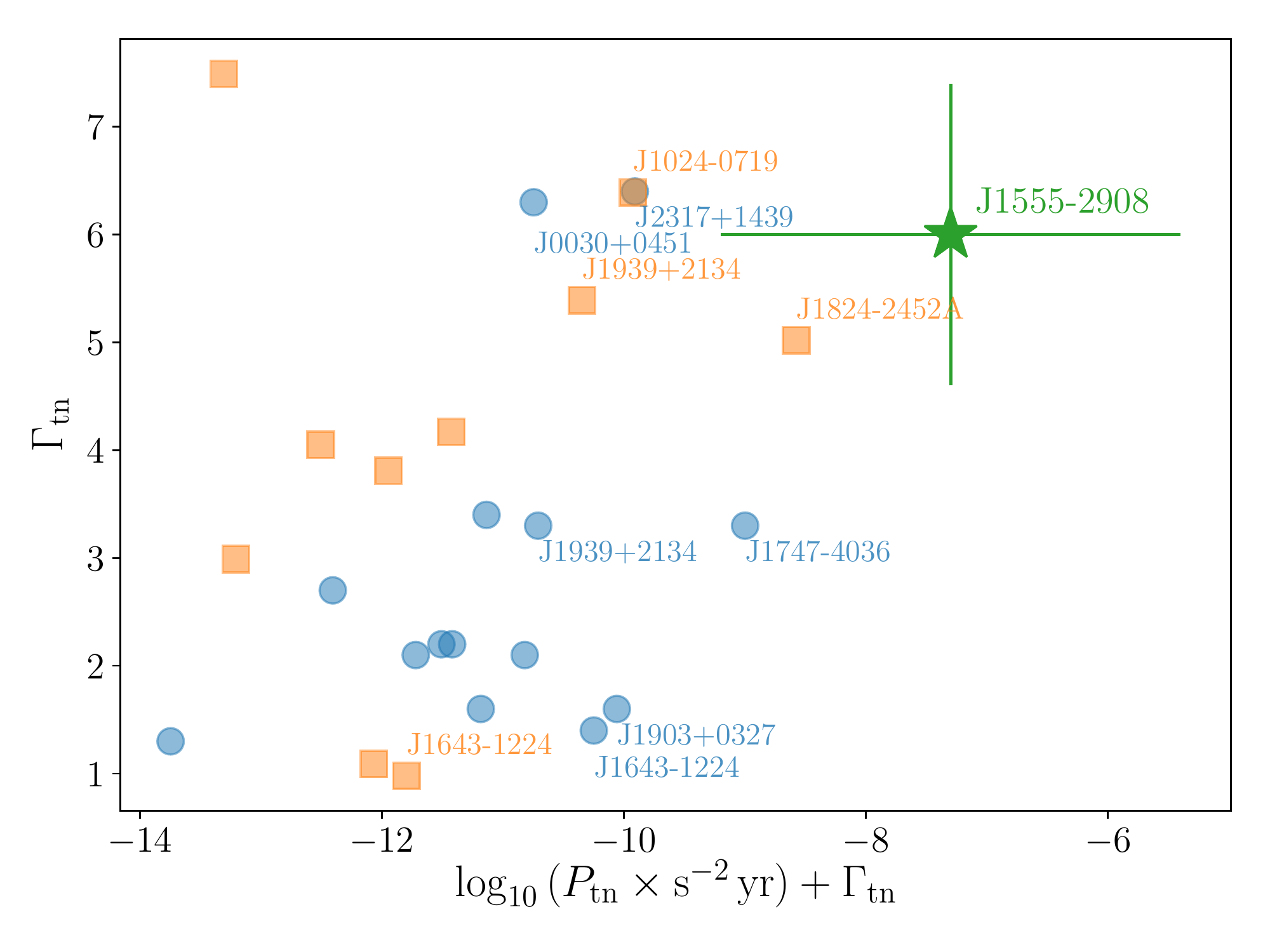}
	\caption{\label{f:j1555_tncomp} Measured values of the timing noise power $P_{\mathrm{tn}}$ and spectral index $\Gamma_{\mathrm{tn}}$ from NANOGrav (blue circles) and the PPTA (orange squares).  The addition of $\Gamma_{\mathrm{tn}}$ to the abscissa converts the power normalization from a frequency of $f=1\,\mathrm{yr^{-1}}$ to $f=0.1\,\mathrm{yr^{-1}}$.  The measured value for \psr{} (this work) is depicted with a green star.  Pulsars with strong timing noise are labeled.}
\end{figure}

The spin-frequency variations of \psr{} (see also Paper~I) are unusually large in amplitude for the timing noise of an \ac{MSP} \citep{kerr2015b}.  Indeed, Figure \ref{f:j1555_tncomp} shows the timing noise spectral properties for two samples of \acp{MSP} as measured by the pulsar timing arrays NANOGrav \citep{Arzoumanian20} and the PPTA \citep{Goncharov21a}, along with our measured values for \psr{}.  It is possible this collection of timing noise measurements is biased by selection for inclusion in a Pulsar Timing Array (PTA).  However, the driving concern for inclusion is radio brightness, and often timing noise emerges only after many years of observation.  Consequently, we expect such a bias to be modest.  The abscissa in Figure \ref{f:j1555_tncomp} is defined in such a way to indicate the timing noise at frequency $f=0.1\,\mathrm{yr}^{-1}$, i.e.\ a time scale comparable to the length of the data set.  \psr{} is a clear outlier, with more than $10$ times the timing noise amplitude of the next strongest pulsar, J1824$-$2452A in the M28 globular cluster, whose timing noise presumably includes contributions from the cluster potential and close passages by other cluster members.  Another noteworthy example is PSR\,J1939$+$2134, an isolated \ac{MSP} considered to be a ``noisy'' timer for which \citet{shannon2013} invoked an asteroid belt surrounding the pulsar to explain the observed timing variations, but which has a timing noise amplitude that is $1000$ times weaker than that of \psr{}.

In this work we have presented an alternative explanation, in which another companion orbits the inner binary system, a hierarchical triple system.  To account for the additional body in the system, we have developed a simple, but sufficient rotational phase model.  In a timing analysis with current data, the results with both models are indistinguishable.  Within the timing range the maximum phase offset between both models is $<1\%$ of one rotation.

It is predictable that the two models would not differ much.  The variations are measurable with the \ac{LAT} data, but the peak-to-peak amplitude of the phase offset is still only $17\%$ of one rotation.  Furthermore, the potential outer companion would have completed roughly one orbit during the \Fermi{} mission which can be well approximated with the six spin parameters of the FX model.

\begin{figure}[t]
	\centering
	\includegraphics[width=\columnwidth]{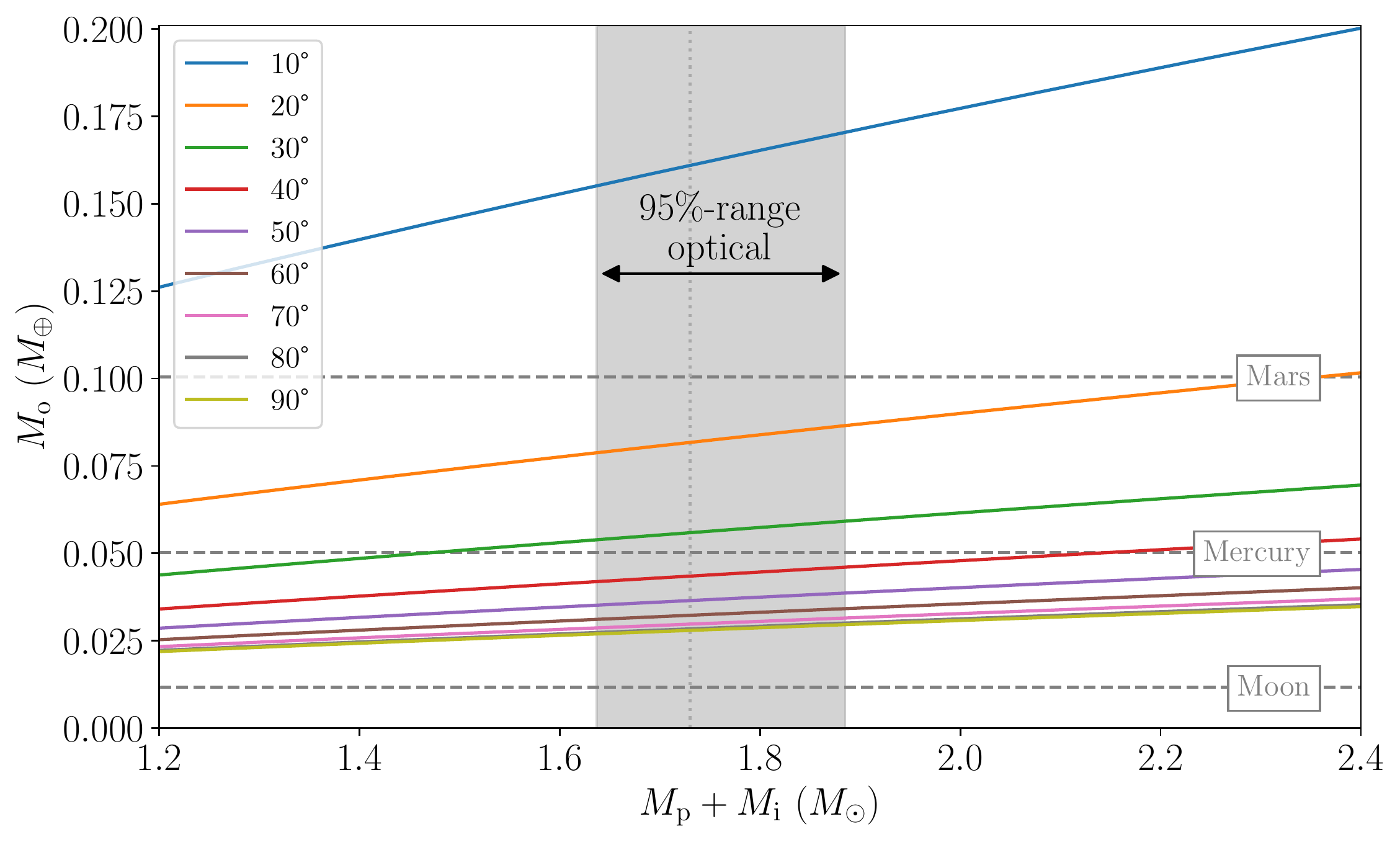}
	\caption{\label{f:j1555_massmass} Mass-mass diagram where lines show the potential outer companion mass for different inclination angles (measured with respect to line-of-sight to Earth) and total masses of the inner binary system.  For comparison, the masses of the Moon, Mercury, and Mars are indicated by the dashed lines.  The dotted line and the gray area mark the most likely value for $M_{\rm p} + M_{\rm i} = 1.73\,M_{\odot}$ and the associated $95\%$-range from optical modeling (Kennedy et al. 2022, submitted).}
\end{figure}

In the following, we will make estimates of the companion mass and the orbital radius, but want to note that these directly depend on the outer orbital period $P_{\rm o}$.  Due to the strong correlations between the orbital parameters (see Section~\ref{s:gamma}), these estimates should be treated with caution and will only give an order of magnitude.

The binary mass function may be used to estimate the mass of the outer companion (see Fig.~\ref{f:j1555_massmass}).  For common pulsar masses $1.4\,M_{\odot} < M_{\rm p} < 2.0\,M_{\odot}$, an inner companion mass $M_{\rm i} = 0.06\,M_{\odot}$ (Kennedy et al. 2022, submitted), and inclination angles $~30\degr-90\degr$, the outer companion would have roughly a Mercury-like mass $0.02\,M_{\oplus} \lesssim M_{\rm o} \lesssim 0.06\,M_{\oplus}$.  To date, this companion would still be listed among the four lowest-mass extrasolar planets\footnote{\href{http://exoplanet.eu/catalog/}{http://exoplanet.eu/catalog/}}.

A comparably low-mass ($\sim 0.015\,M_{\oplus}$) companion has also been found for a planet orbiting PSR\,B1257$+$12 \citep{Wolszczan1994Planet}, the pulsar thanks to which, $30$ years ago, the first extrasolar planetary companions have been identified \citep{Wolszczan1992Planet}.  However, unlike the periods of the planets orbiting PSR\,B1257$+$12 with tens to hundreds of days, the probable companion of \psr{} takes considerably longer with $P_{\rm o} \sim 4{,}500$ days.

The putative triple system of \psr{} is extremely hierarchical.  The pulsar's mass is $\sim 28$ times larger than the inner companion's mass, and even $\sim 10^7$ times larger than the outer companion's mass.  The ratio between outer and inner orbital period is $P_{\rm o}/P_{\rm i} \approx 20{,}000$.  Using Kepler's third equation, the masses from optical modeling, and the timing parameters one can estimate the semimajor axis of the inner companion's orbit as $\sim 4.2$\,lt-s and the outer companion's orbit as $\sim 3{,}200$\,lt-s.  Systems with such extreme hierarchical properties are typically assumed to be stable on long timescales \citep{Georgakarakos2008+Stability}.

Other \acp{HTS} which include a pulsar are known.  PSR\,J0337$+$1715 is in an orbit with two white dwarf companions \citep{ransom2014}.  Carefully studying this system led to one of the most stringent tests of General Relativity's predicted universality of free fall \citep{archibald2018,Voisin2020+J0337}.  A binary system consisting of the \ac{MSP} PSR\,B1620$-$20 and a white dwarf is orbited by a planetary companion of $\sim 2.5\,M_{\rm Jupiter}$ \citep{Thorsett1993+PSRwithPlanet,Sigurdsson2003+PSRwithPlanet}.

An open question is how such a system could have formed.  A pulsar planet can form in various ways \citep[see, e.g.,][]{Podsiadlowski1993+Formation}.  Since \psr{} is not in a dense Globular Cluster, it seems unlikely that the triple was formed in an exchange interaction between systems as is suggested for PSR\,B1620$-$20 \citep{Sigurdsson2003+PSRwithPlanet}.  Another option could be that the inner binary captured a free-floating planet \citep[see, e.g.,][]{MiretRoig2021+FreePlanets}.  On the other hand, the planet might have formed from a circumbinary debris disk \citep[see, e.g.,][]{Pelupessy2013+formation}, in which case it would be expected that the orbits are coplanar.

Only to emphasize the power of pulsar timing with \ac{LAT} data, we want to note that if \psr{} has an additional planetary-mass companion, then it was discovered because over the course of $100$ billion pulsar rotations, the phase drifted by $17\%$ of one single rotation.

So, is \psr{} in a hierarchical triple system?  We do not know, yet.  With current data, both of the timing models that are presented here track the pulsar's rotation equally well.  However, the ongoing \Fermi{} mission will continue to collect data, and eventually the nature of this system will be resolved.

%%%%%%%%%%%%%%%%%%%%%%%%%%%%%%%%%%%%%%%%%%%
\acknowledgments{}

This work was supported by the Max-Planck-Gesellschaft~(MPG). C.J.C. acknowledges support from the ERC under the European Union's Horizon 2020 research and innovation programme (grant agreement No. 715051; Spiders). Portions of this work performed at NRL were supported by NASA.

The \textit{Fermi} LAT Collaboration acknowledges generous ongoing support from a number of agencies and institutes that have supported both the development and the operation of the LAT as well as scientific data analysis.  These include the National Aeronautics and Space Administration and the Department of Energy in the United States, the Commissariat \`a l'Energie Atomique and the Centre National de la Recherche Scientifique/Institut National de Physique Nucl\'eaire et de Physique des Particules in France, the Agenzia Spaziale Italiana and the Istituto Nazionale di Fisica Nucleare in Italy, the Ministry of Education, Culture, Sports, Science and Technology (MEXT), High Energy Accelerator Research Organization (KEK) and Japan Aerospace Exploration Agency (JAXA) in Japan, and the K.~A.~Wallenberg Foundation, the Swedish Research Council and the Swedish National Space Board in Sweden.

Additional support for science analysis during the operations phase is gratefully acknowledged from the Istituto Nazionale di Astrofisica in Italy and the Centre National d'\'Etudes Spatiales in France.  This work performed in part under DOE Contract DE-AC02-76SF00515.

%\facilities{}

%\software{}

%%%%%%%%%%%%%%%%%%%%%%%%%%%%%%%%%%%%%%%%%%%
\bibliographystyle{aasjournal}

\bibliography{library_j1555}

\end{document}